\documentclass{aastex}

\shorttitle{PSR B0656 proper motion}
\shortauthors{R.P. Mignani et al.}

\begin{document}

\title{HST revisits the proper motion of PSR B0656+14\footnote{Based
on observations with  the NASA/ESA Hubble  Space
Telescope, obtained at the Space Telescope Science Institute, which is
operated by AURA, Inc., under NASA contract NAS 5-26555.}}

\author{R.P. Mignani }
\affil{STECF-ESO, Karl Schwarzschild Str.2, D8574O Garching b. Munchen}
\email{rmignani@eso.org}

\author{A. De Luca,P.A. Caraveo\altaffilmark{1}}
\affil{IFC-CNR, Via Bassini 15, I-20133 Milan}

\altaffiltext{1}{Istituto Astronomico, Via Lancisi 29, I-00161 Rome}

\begin{abstract}

We report the  first   successful measurement  (with HST/WFPC2)  of  a
significant proper motion for a m$_{\rm V}\sim$ 25 object which we had
proposed as the optical counterpart of  the radio pulsar PSR B0656+14.
The yearly  displacement of $43\pm  2$ mas is  highly significant and,
while in agreement with early, rougher, radio measurements, stands out
for  its vastly improved  accuracy.  Together  with  the  report of  a
possible optical pulsation \citep{she98}, we regard this result as the
final proof of the optical identification of PSR B0656+14. 

\end{abstract}


\section{Introduction}

PSR B0656+14  is  a middle-aged  ($\sim100,000$  yrs) isolated neutron
star, detected as a radio (385 ms) as well as an X-ray pulsar by ROSAT
\citep{fin92},  which confirmed the  X-ray   pulsation hinted in   the
discovery data of  the  EINSTEIN satellite \citep{cor89}.   A possible
identification  of    the  pulsar in    gamma-rays  was  also  claimed
\citep{ram96} but so  far only a  marginal evidence  of pulsations was
found  in the GRO/EGRET  data.  Although  dispersion measure estimates
put PSR B0656+14  at $\sim$ 760 pc, its  true distance  remains highly
uncertain.  Indeed, assuming for the  neutron star a canonical  radius
of 10 km, blackbody  fits to the soft thermal  component of its  X-ray
spectrum \citep{fin92}  impose a significant  downward revision of the
nominal distance.   PSR B0656+14  could thus  join  the thin  crowd of
pulsars detected within few hundred pc of the Sun.  Since PSR B0656+14
is  inside the putative supernova remnant  known as  Monogem Ring, the
measurement  of its proper  motion  was a  natural way to  investigate
their   possible association. Thus,  3   radio  images of  the  field,
covering  roughly 4 years, were   collected by \citet{tho94} from  the
NRAO  Very  Large     Array  yielding  a  proper    motion   value  of
$\mu_{\alpha}=64\pm11$  mas/yr; $\mu_{\delta}=-28\pm4$   mas/yr    and
ruling  out  the possible association.  The   proper  motion value was
later revised  by \citet{pav96}  in $\mu_{\alpha}=73\pm20$  mas/yr and
$\mu_{\delta}=-26\pm13$    mas/yr  (PA=$109^{\circ}\pm   10^{\circ}$),
which, although rather  uncertain,  shall   be  taken as the     radio
``reference'' value.    In the optical,  a   likely counterpart to PSR
B0656+14  ( m$_{\rm V}\sim$ 25)  was proposed by  \citet{car94} on the
basis of its  positional coincidence with the  radio  pulsar.  For the
possible range of  the pulsar distance  ($800\div250$ pc), the optical
emission  of the candidate counterpart was  found to be well above the
predictions based  on    both the magnetospheric     emission model of
\citet{pac87}  and the   extrapolation   of the soft  X-ray  blackbody
emission \citep{fin92}.   Indeed,  multicolor photometry,  reported by
\citet{pav97}, gave  support  to  this  identification by    showing a
composite  spectral shape,      which   could be interpreted   as    a
magnetospheric component superimposed to a thermal one.  To assess the
reliability  of  the proposed   optical  identification, \citet{mig97}
tried to use proper  motion as a  distinctive character of the pulsar.
The    original ESO/3.6m and   NTT  observations of \citet{car94} were
compared to a newly acquired HST/WFPC2  one to search for the object's
displacement.    At  that  time,    the  limited positional   accuracy
achievable with  the ground   based  images hampered  the  astrometric
measurements  and   no  conclusive result  could   be  obtained.  Soon
thereafter,  optical  pulsations  at   the  radio/X-ray period    were
detected,   at  a 4  $\sigma$ level,   from   the proposed counterpart
\citep{she98}, thus   apparently  closing   the identification  issue.
However, the limited significance of this  detection calls for a clear
confirmation  through   new and independent   observations.   Thus, we
started a successful  programme of HST observations  aimed  at a solid
detection of the candidate's proper  motion.  In this paper,  we
report  the outcome   of  this programme:  the   observations and data
reduction  are  described  in    section~\ref{obsdatared},  the results   are
discussed in  section~\ref{res}, while the comparison  between optical
and radio proper motion  values is given in  section~\ref{optvsr}. The
implications of our measurement are summarized in section~\ref{concl}.

\section{Observations and data reduction} \label{obsdatared} 

\placefigure{fig1}

The   HST   observations were performed  on    January  18th 1996, see
\citet{mig97}, and  on  January  14th   2000,  just after  the   third
refurbishing  mission.  In   both   cases,  to  maximize the   angular
resolution,  the  target was  centered in the   Planetary Camera (PC),
which has a pixel  size of 45.5 mas.   The images were taken  with the
555W filter ($\lambda=5252 \AA$;   $\Delta \lambda= 1222.5 \AA$).  The
pulsar was observed for three  orbits in 1996  and  for two orbits  in
2000, which correspond to  total exposure times of  6200 s and 4400 s,
respectively. The observations were performed  with the same telescope
roll angle chosen  in such a  way that the PC  X,Y  axis were oriented
along right ascension and declination. 

After standard pipeline processing, the images  have been cleaned from
cosmic ray hits  by   combining repeated  exposures through  a  median
filter.    Following    the  standard  astrometric     approach,  e.g.
\citet{car96}, the frames registration has been performed by computing
the pixel coordinate transformation between a  common set of reference
objects identified in the two images and chosen to be neither extended
objects nor relatively bright and   saturated stars.  The 6  reference
objects that  satisfy our criteria,  together  with the pulsar optical
counterpart, are  identifiable  in  the  January  2000  image shown in
Fig.1.  Their pixel coordinates  have  then been  computed in  the two
frames by  2-D gaussian fitting   of PSF profiles.  The procedure  was
repeated   for  different  widths  of  the   centering  box  until the
coordinate  values were shown to be  stable. Finally, the centroids of
the reference    objects have been   determined  with an   accuracy of
$\sim0.05$ px for  both epochs.  The  coordinates of the  PSR B0656+14
counterpart were computed  following the  same approach.  Of   course,
owing to its lower S/N, its  centering error was found slightly higher
and  more   dependent  on the  size  of  the    centering box.   To be
conservative, we have rounded  up the associated  error to 0.1 px. The
quoted  uncertainties also  account   for  possible centering   errors
induced by the subtle exposure-to- exposure  jitter ($<0.04$ pixels in
both frames).  All the measured  centroids were subsequently corrected
for the  effects   of the  PC  geometric distorsion   \citep{gil95} by
applying the ``metric'' task  of  the STSDAS package.   The coordinate
transformation  for the  2000 vs 1996  frame  registration was finally
computed through a linear   fit  (rotation plus offset) between    the
second and the  first  reference  grid.  The  computed  transformation
turned out to be accurate to within 0.07 and 0.2 px  along the X and Y
axis, respectively.  We  note that the   errors on  the trasformation,
certainly higher than the ones obtained e.g.  by the \citet{del00a} in
the case of the Vela pulsar, are  independent of the alghoritm used to
fit the centroids of  the reference stars. This  effect is probably to
be  ascribed to   the  much lower number  of  reference  stars used to
compute  the coordinate  transformation  as well  as to their relative
distribution in  the  frame. Indeed,  the error   along Ra,  where the
reference stars are distributed on a larger area, is much smaller than
the one in Dec, where the reference star are much closer to each other
(see Fig.1).

\section{Results} \label{res}

Having secured  a single reference frame, we  can compare the relative
positions of  the pulsar optical  counterpart at the  two epochs.  The
total  offset  turns  out  to be   $3.76\pm0.16$ px, where  the  error
includes  all the  uncertainties  of the  astrometric steps (exposures
alignement, object centering and frame registration). The displacement
is obviously very significant and it represents  a convincing proof of
the object's motion.  After  conversion from pixel to sky coordinates,
the measured  offset  translates   into  an angular  displacement   of
$\sim172\pm8$ mas over 4 years,  corresponding to  a proper motion  in
the plane of the sky of $\sim43\pm2$ mas/yr.  We note that this result
can  not be improved by using  the 1989/1991  images of \citet{car94},
which, although providing a $\approx 3$ times  longer time span, would
introduce an  error larger than 0\farcs2 on the corresponding
target   position,  worsening significantly  affecting  the  accuracy   on  the pulsar
displacement.  We note that the measured proper motion turns out to be
almost     totally in   right ascension    with   the two   components
$\mu_{\alpha}=42.7\pm2$     mas/yr,     $\mu_{\delta}=-2.1\pm3$ mas/yr
defining  a position  angle of $93^{\circ}\pm   4^{\circ}$.  Since our
WFPC2 observations  have been taken virtually on  the same  day of the
year  (january  18 and 14,  respectively),  we  can exclude  that  the
measured displacement is affected by the object parallax, which, for a
distance  of e.g.  300 pc, would  be $\sim3$ mas  (almost all in right
ascension).    Thus,  the measured  displacement is   due to a genuine
proper motion of the source. 

\subsection{Optical vs. Radio} \label{optvsr}

We can now compare  our optical proper motion ($\mu_{\alpha}=42.7\pm2$
mas/yr, $\mu_{\delta}=-2.1\pm3$ mas/yr)  to the  revised radio one  of
\citet{pav96}        i.e.   $\mu_{\alpha}=73\pm20$   mas/yr        and
$\mu_{\delta}=-26 \pm13$ mas/yr.  Although certainly compatible within
their  quoted errors,  the optical  result  appears much more accurate
than the  radio one.  Since  both measurements span  about 4 years and
use   a comparable  number  of   reference objects \citep{tho94},  the
difference  in accuracy is  ascribable  both to systematic errors  that, in
spite of the strengh of the signal from  the radio pulsar, pleague the
VLA data (see discussion in Pavlov, Stringfellow and Cordova 1996) and
to the higher angular resolution of the WFPC2. \\
On the other hand, the reliability of HST astrometry has been recently
assessed by DeLuca et al. (2000b) through  comparing independent couples of WFPC2
observations  of the Vela and  Geminga pulsars, which yielded, for each
object, fully consistent proper motion measurements.


\section{Conclusions} \label{concl}

Taking advantage  of   the outstanding performances   of the Planetary
Camera on HST, we have secured a firm measurement of the proper motion
of the candidate optical counterpart of  PSR B0656+14, which moves at a
yearly  rate of  $43\pm2$  mas.  Even   in the  absence of  any  other
supporting evidence, the presence of such a proper motion for a 25 mag
object, coupled  with a distance in  the range $250\div800$  pc, would
imply  a transverse velocity anywhere between  50  to 160 km/sec. Such
values are typical of just one  class of astronomical object: isolated
neutron stars.  This was the case of  Geminga, when the measurement of
the  proper motion of the  optical  counterpart \citep{big93} clinched
the   identification of  the   faint G'' with   one   of the brightest
gamma-ray  source in  the  sky   \citep{big96}.  Although similar   to
Geminga in  many ways, PSR B0656+14 does  not offer  such a challenging
situation: we  face a bona fide radio  pulsar,  with a  radio
proper motion and a  promising optical counterpart, the identification
of which is supported both by  multicolor photometry \citep{pav97} and
timing \citep{she98}. However, although both methods appear suggestive
and definetely worth pursuing, the burden of the proof of PSR B0656+14
optical   identification should rest  on a  firmer ground.  Our  highly
significant measurement of a  proper motion of the optical counterpart
provides just such a straighforward evidence. 

\acknowledgements We   wish  to thank Giovanni  F.  Bignami   for many
stimulating discussions.   ADL thanks  the  Space Telescope   European
Coordinating Facility  for hospitality and  support.  This research is
supported by the Italian Space Agency (ASI).

\figcaption[fig1.ps]{HST/WFPC2  image   of   the  PSR B0656+14   field
obtained in Jan. 2000  through the  555W  filter for a  total exposure
time of 4400  s.   The image, showing  the  PC chip only,  is oriented
along  right ascension and declination (north  to the top, east to the
left).  The full   image f.o.v.  is  $36\times36$ arcsec.   The pulsar
counterpart is  labelled together with the  reference stars (A-F) used
for  relative  astrometry measurements  (see  text).   The solid arrow
marks    the   direction     of    the  computed      optical   proper
motion.\label{fig1}}

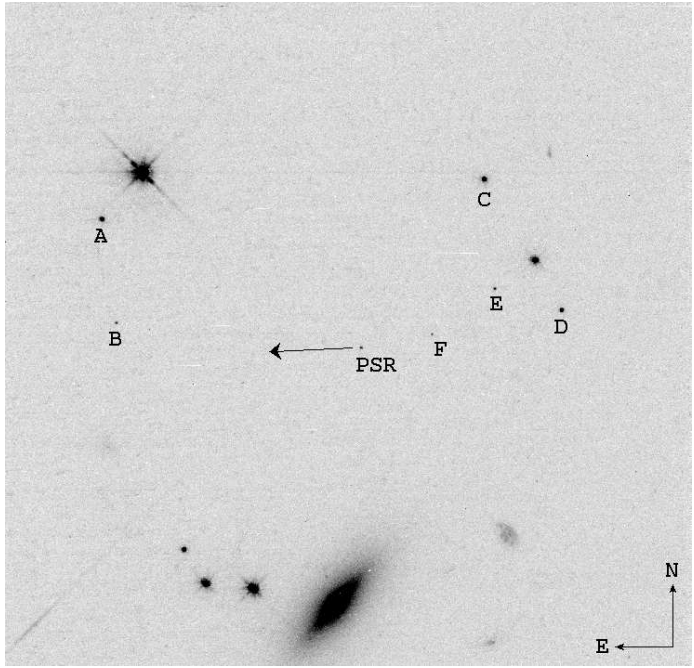
\end{document}